\documentclass[final]{article}

\usepackage{mathptmx}       % selects Times Roman as basic font
\usepackage{helvet}         % selects Helvetica as sans-serif font
\usepackage{courier}        % selects Courier as typewriter font
\usepackage{type1cm}        % activate if the above 3 fonts are
\usepackage{makeidx}         % allows index generation
\usepackage{graphicx}        % standard LaTeX graphics tool
\usepackage{multicol}        % used for the two-column index
\usepackage[bottom]{footmisc}% places footnotes at page bottom
\usepackage{amsmath}
\usepackage{amsfonts}
\usepackage{amssymb}
\usepackage{texnames}
\usepackage{url}

\usepackage{xspace}
\usepackage[]{graphicx}
\graphicspath{{image/}}

\usepackage{listings}
\usepackage{color}
\definecolor{middlegray}{rgb}{0.5,0.5,0.5}
\definecolor{lightgray}{rgb}{0.9,0.9,0.9}
\definecolor{orange}{rgb}{0.8,0.3,0.3}
\definecolor{yac}{rgb}{0.6,0.6,0.1}
\definecolor{javared}{rgb}{0.404,0,0.023}
\definecolor{javagreen}{rgb}{0.004,0.161,0.02}

 \lstdefinelanguage{XQuery}[]{}{
        morekeywords={for, let, as, function, declare, if, then, else, insert, node, into, return},
        sensitive=false,
        morecomment=[s]{(:}{:)},
        morestring=[b]',
        morestring=[b]"        
}
 \lstdefinelanguage{XMLSchema}[]{XML}{
        morekeywords={schema, complexType,sequence, element}      
}
 \lstdefinelanguage{XSL}[]{XML}{
        morekeywords={xsl:stylesheet,xsl:output,xsl:template}      
}
 \lstdefinelanguage{DGDXML}[]{XML}{
        morekeywords={Model, Description, Title, Author, Text, Cite, Keyword, Reference, Field, Date, MediaObject, File}      
}

\newcommand\dgdgallery{\textsc{DGD Gallery}\xspace}

\makeindex             % used for the subject index
\begin{document}

%opening
\title{DGD Gallery: Storage, sharing, and publication of digital research data} 
\author{Michael Joswig \and Milan Mehner \and Stefan Sechelmann \and Jan Techter \and Alexander I. Bobenko}
%\authorrunning{M. Joswig \and M. Mehner \and S. Sechelmann \and J. Techter \and A.I. Bobenko}

\maketitle

\section{Introduction}
Software produces data.
Mathematical software produces scientific data, and this is often worth keeping.
One reason for this can be the vast amount of CPU time spent on a specific experiment.
Another reason can be that the output is obtained only via a complex interaction process between the software and its user.
That latter situation is typical in mathematical visualization, where producing a satisfying or even beautiful picture of a geometric object is a form of art.
The purpose of this text is to describe a new project, called the ``Discretization in Geometry and Dynamics Gallery'', or \dgdgallery for short, whose goal is to store geometric data and to make it publicly available.
The URL of the web-site is
\begin{quote}
 \url{http://gallery.discretization.de}
\end{quote}
Today it is safe to say that finally mathematical software has reached every branch of mathematics.
While computers have played a role in mathematical applications for a long time, it took considerably longer for software to be appreciated fully in parts of mathematics traditionally considered as ``pure''.
The array of tools at our fingertips now includes solvers for linear programs and partial differential equations, but also software for dealing with real algebraic sets \cite{bertini} or delicate constructions in sheaf theory \cite{BarakatLange-Hegermann:1409.2028}.
To get an idea of how rich the mathematical software landscape has become, see, e.g.,~\cite{ICMS2014}.
The success of each single software system raises the question of how the respective data produced should be stored.
With an increasing number of relevant mathematical results relying on non-trivial computations in an essential way (see, e.g, the \texttt{Flyspeck} project \cite{flyspeck:1501.02155}) it becomes more and more crucial to publish such results in a way such that they can be scrutinized (and used) by the mathematical community.

The mathematical data we have in mind for the \dgdgallery are the geometric objects that occur naturally on the border between differential geometry and geometric combinatorics.
This includes various classes of surfaces (embedded or immersed) in
$3$-space, convex polytopes and polyhedral fans of various dimensions, circle patterns, and many more.
Yet, we believe that several of our design decisions and architecture ingredients will be useful for other collections of mathematical data.
Key features include the following:
\begin{itemize}
\item structured storage of research data,
\item review process for increased reliability,
\item migration process for sustainability,
\item licensing scheme.
\end{itemize}
To further stress the relevance of our endeavor, it is worth noting that scientific funding agencies have begun to add requirements concerning the preservation of scientific data to their regulations.
For instance, in a recent announcement \cite{DFG} of Deutsche Forschungsgemeinschaft (DFG) says:\footnote{Translated from German.}
\begin{quote}
  The documentation of research data according to standards depending on the subject and their long-term archival are relevant for controlling the quality of scientific work.
  Further, these data are the basic requirements for the subsequent use of research results.
\end{quote}
% Die Dokumentation von Forschungsdaten nach fachspezifischen Standards und ihre langfristige Archivierung sind daher nicht nur bedeutsam für die Qualitätssicherung wissenschaftlicher Arbeit, sondern auch eine grundlegende Voraussetzung für die Nachnutzbarkeit von Forschungsergebnissen.

The \dgdgallery evolved as a project within the DFG Collaborative Research Center SFB/TR 109 ``Discretization in Geometry and Dynamics''.
Its usage is currently restricted to the members of the center. However, it is intended that future versions allow other researchers to contribute their work, too.
%The content of the gallery is stored on the servers of the research center that are part of the infrastructure of TU Berlin. 
%In this way we ensure that the service will be reliably available during and after the funding period of the CRC.

The paper is organized as follows.
First we compare our design to existing collections of geometric data (Section~\ref{sec:previous-work}).
Then, in Section~\ref{sec:examples}, we exhibit some examples already published on the gallery.
This should give a good idea of what kind of collection we have in mind.
At the same time this also shows some of the technical features and capabilities.
The core is Section~\ref{sec:architecture}, where we elaborate on the architecture and the design decisions.
The key concept is the \emph{model}, which is our technical realization of a geometric object.
Some aspects of the implementation are covered in Section~\ref{sec:implementation}.
For instance, we explain how we use the XML document database \textsc{BaseX}~\cite{basex-website} and meet current standards of web technology.

\section{Comparison With Previous Work}
\label{sec:previous-work}
To store geometric data digitally and make it accessible through a web-site is clearly not a new idea.
On the contrary, since the early days of the Internet people have set up numerous web-sites with all kinds of information on geometric objects, e.g.: The Geometry Center's ``Geometry Reference Archive'' \cite{geometry-center-website}, ``The Scientific Graphics Project''of MSRI \cite{msri-sgp-webpage}, or David Eppstein's ``Geometry Junkyard'' \cite{eppstein-website}, to name a few prominent examples.
Clearly, all of the above still contain lots of interesting information.
However, there are some shortcomings.
In the case of the archive of The Geometry Center we have a static collection of data that will not see any updates or additions.
Yet there is the advantage that all data is available from one source, and so it cannot degenerate over time (except for eventually outdated file formats).
Not so with the ``Geometry Junkyard''.
This is a collection of links to other interesting resources on the web.
Many of the links are dead already.
This is mainly due to a discontinued provider service or simply a change of position of the person who provided the data.
The ``Scientific Graphics Project'' is mainly a collection of surfaces and differential geometry related publications.
The \dgdgallery wants to cover geometric objects from a much wider collection.

A more recent project is the "GeometrieWerkstatt"~\cite{geometrie-werkstatt-webpage} maintained by a group of geometers at T{\"u}bingen University.
It contains visualizations of mostly smooth constant mean curvature surfaces. Surfaces are visualized using
videos, images, and interactive 3D viewers. 
The main difference to the \dgdgallery is that there is no geometric data that can be accessed via the web-site.
On the other hand, how to provide the data for smooth surfaces is far from obvious and cannot be separated from 
the mathematical methods. 
For the \dgdgallery we propose to include a discretization of a smooth surface in a reasonable resolution.

"{IMAGINARY -- open mathematics}" is a platform~\cite{imaginary-website} which has a strong educational focus.  It
features images and mathematical software for a broad audience such as exhibitions, high school education, and museums.
As an essential feature, IMAGINARY is open for the public to contribute material by cross-linking to other web-sites.  In
this way it works like a collection of collections.

The focus of the SymbolicData project \cite{symbolicdata-website} is on developing concepts and tools for profiling, testing and benchmarking Computer Algebra Software.
This includes storing scientific data from various sources, but visualization does not play a role.

The project that is most similar in spirit to our \dgdgallery is ``Electronic Geometry Models'' \cite{eg-models, eg-models-website}, which is a refereed online journal for digital geometry models on the web.
It features XML file formats, visualization separated from descriptions, and a reviewing process.
All of these are also implemented in the \dgdgallery.

However, our technical realization substantially differs from ``Electronic Geometry Models''.
The \dgdgallery employs modern web technology for the user interface and a standard data base implementation for storing.
One advantage of this is the possibility to work in teams.
Each team member can contribute to a model if he/she is a registered user with the suitable permissions.
Permissions can be granted by owners of content; for details see Figure~\ref{fig:permissions} below.
Moreover, the entire work flow from the submission, through reviewing and revising, to the final publication has one consistent setup through a common front end.
Most importantly, the overall design is highly modularized.
For instance, the \dgdgallery features a variety of media renderers with different visualization strengths to accommodate for heterogeneous hard- and software environments at the users' end.
This is also relevant for being able to preserve the data over a long period.

Another difference to ``Electronic Geometry Models'' is that the \dgdgallery aims at a broader outreach and therefore seeks to include more models of purely educational value.
This results in a different set of criteria for accepting a model for publication.
Moreover, the \dgdgallery allows for changes to a model after publication.

\section{Examples}
\label{sec:examples}

In this section we present some selected models from the early contributions to the \dgdgallery.
They are intended as guidelines and inspiration for future models to be submitted.

\subsection{Discrete S--Conical Catenoid and Helicoid}
\label{ex:catenoid}
Authors: Alexander Bobenko, Tim Hoffmann, Benno K\"onig, and Stefan Sechelmann

\begin{quote}
\texttt{\small http://gallery.discretization.de/models/sc-catenoid}
\end{quote}

\noindent
This model shows discrete s-conical versions of the catenoid and the helicoid, which are classical minimal surfaces~\cite{BHKS15}.
The smooth versions are among the first classical minimal surfaces ever investigated.
Their s-conical counterparts are quadrilateral polyhedral surfaces with the property that at each vertex the adjacent faces are tangent to a cone of revolution.
The theory of these discrete minimal surfaces is closely related to the theory of orthogonal circle patterns and Koebe polyhedra; see Section~\ref{subsec:koebe} below.
Its features and constructions are similar to the theory of s-isothermic surfaces.
% investigated in~\cite{BobHofSpr06}.
%The dual surface, which exists for smooth minimal surfaces, can be realized from a Koebe polyhedron in the discrete setting.
A minimal surface is (Christoffel) dual to its Gauss map.
This property is preserved in the discrete setup, and so discrete minimal surfaces are constructed from Koebe polyhedra.
The associate family of minimal surfaces is contained in the discrete theory as well.

\begin{figure}[h]
\centering
 \includegraphics[width=0.9\textwidth]{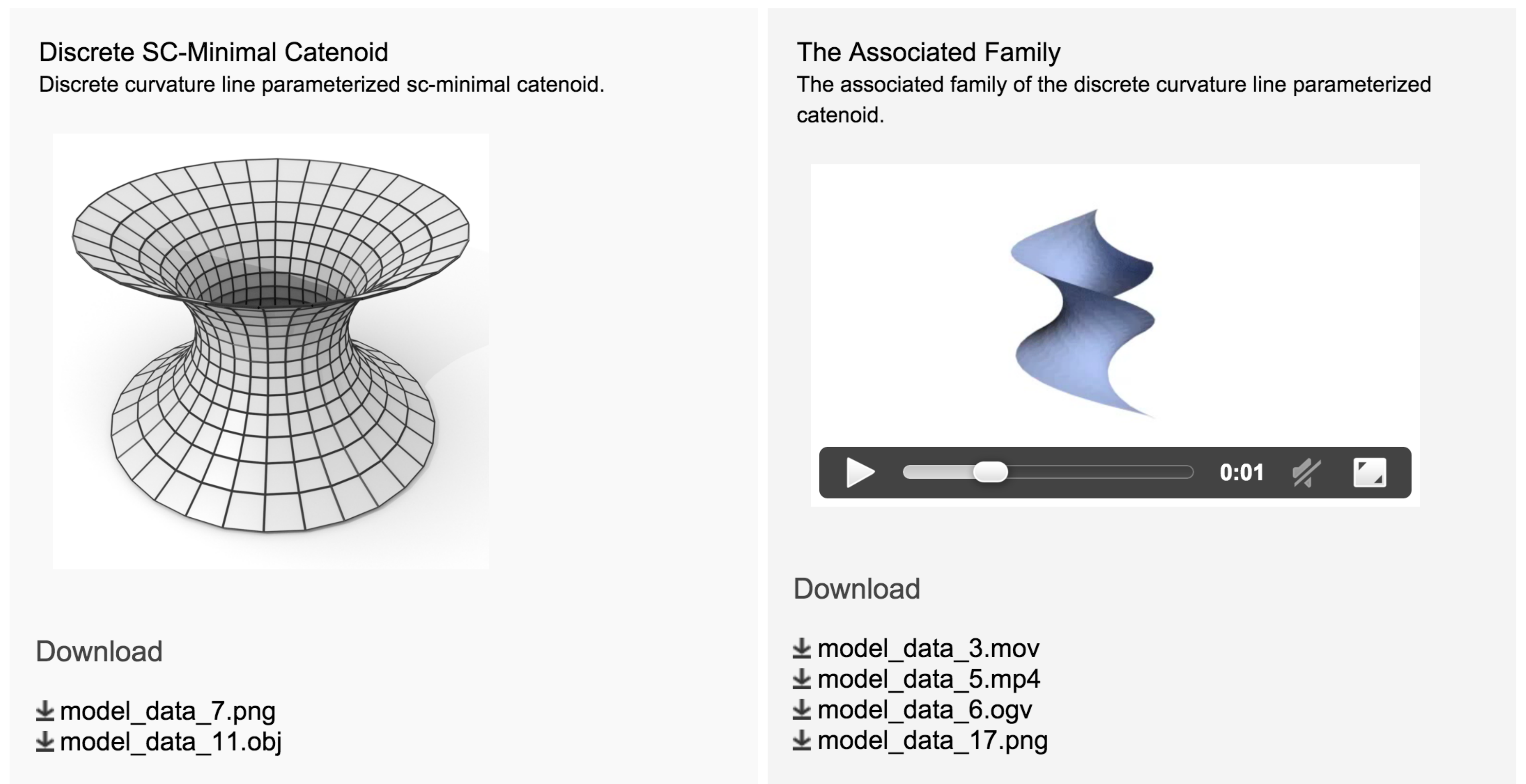}
 \caption{Screenshot of two media objects contained in the "Discrete S-Conical Catenoid and Helicoid" model as presented by a modern web browser.
 Left: Discrete s-conical catenoid. Right: Associate family animation between s-conical catenoid and its conjugate, the discrete helicoid.}
\end{figure}

The model features images of the catenoid and helicoid using representations with discrete curvature line parameterizations as well as discrete asymptotic line parameterizations in the associate family.
It contains a video with an animation of the associate family animating the angle parameter.
Geometric data is given as OBJ files and corresponding preview images.

\goodbreak % let's try with a bit less force; we'll make it work in the end
\subsection{$z^a$ Circle Pattern}

Authors: Jan Techter and J\"urgen Richter-Gebert

\begin{quote}
\texttt{\small http://gallery.discretization.de/models/zalpha\_circle\_pattern}
\end{quote}

\noindent
The representation of discrete holomorphic functions by 
%circle packings goes back to Thurston~\cite{T85}, while the special case of orthogonal 
circle patterns with square-grid combinatorics was first studied by Schramm~\cite{S97}.

This model shows the Schramm type circle pattern corresponding to the holomorphic map $z \mapsto z^a$ for $0 < a < 2$ in the first quadrant of the complex plane.
Taking the centers and intersections of the circles as complex fields on the first quadrant of $\mathbb{Z}^2$, the discrete map was introduced in \cite{B99} as a special isomonodromic solution of the cross-ratio equation (cross-ratio equal to $-1$ on each elementary quadrilateral).
%by Bobenko.
%, and further developed in~\cite{AB00}.
%The asymptotic behavior conjectured in~\cite{AB00} has been proven 
%in~\cite{BI14}
%by Bobenko and Its using the Riemann-Hilbert approach.
The numerics of these discrete maps is studied in~\cite{bornemann2015}.

The model features an interactive Cinderella \cite{cinderella-website} application where the user can adjust the exponent $a$ and the number of circles, see also Section~\ref{sec:frontend}.

\begin{figure}[h]
\centering
 \includegraphics[width=\textwidth]{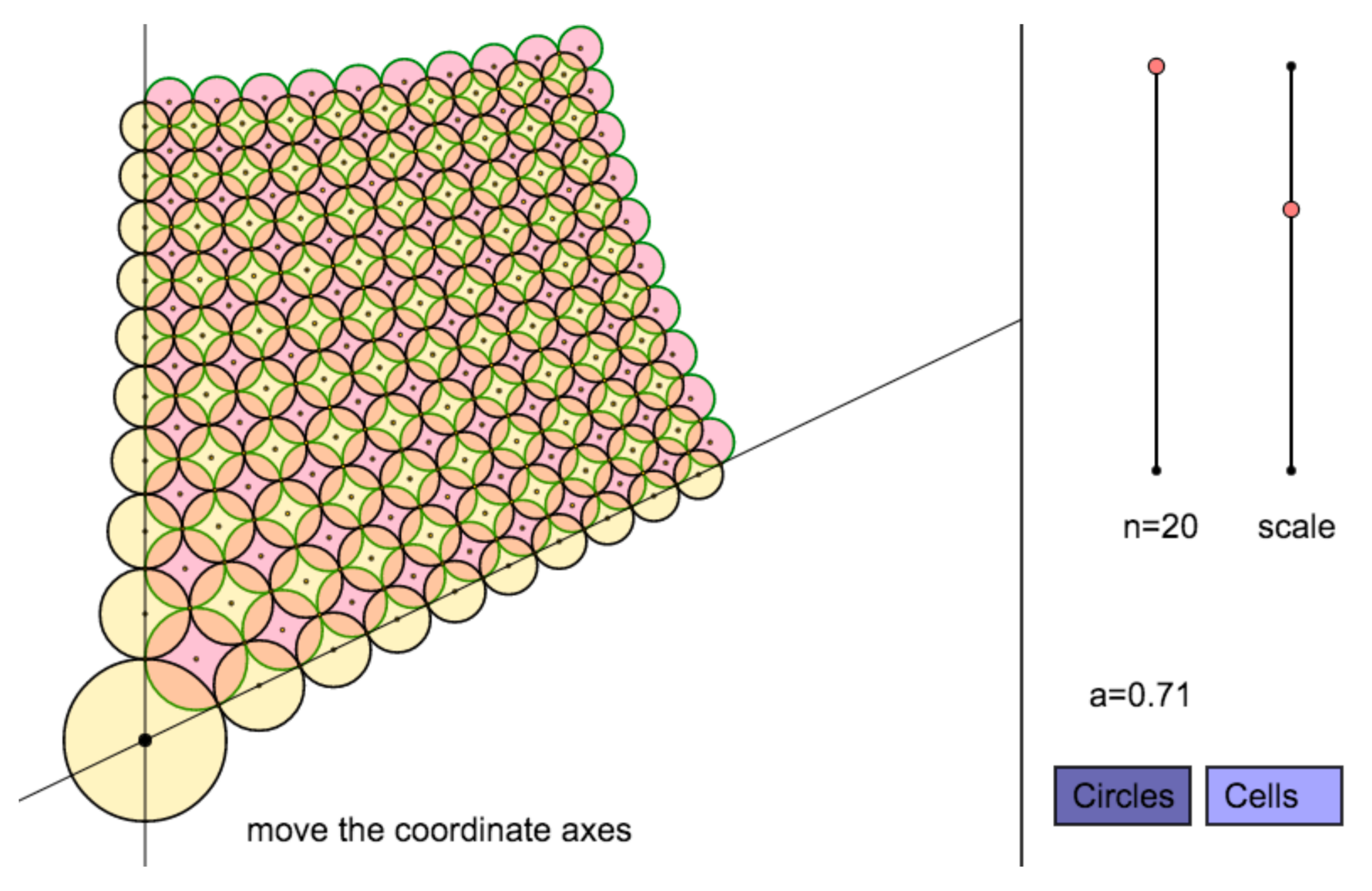}
 \caption{
 Screenshot of the interactive element of the "$z^a$ Circle Pattern" model. 
 A user can adjust the number of circles in a row as well as the overall scale of the drawing.
 The angle $\alpha$ is entered by moving the axes with the mouse.
 }
\end{figure}

\break 
\subsection{Koebe Polyhedra}
\label{subsec:koebe}

Author: Stefan Sechelmann

\begin{quote}
\texttt{\small http://gallery.discretization.de/models/koebe\_polyhedra}
\end{quote}

\noindent
A \emph{Koebe polyhedron} is a $3$-dimensional convex polytope whose edges are tangent to the unit sphere.
Koebe polyhedra have a strong connection to the theory of circle patterns, see \cite{Bobenko04variationalprinciples}.
The theory of discrete minimal surfaces of s-isothermic and s-conical type is based on Koebe polyhedra.
%, see~\cite{BobHofSpr06}.
Each combinatorial type of $3$-polytope admits a representation as Koebe polyhedron, which is unique up to M\"obius transformation.

\begin{figure}[h]
\centering
\includegraphics[width=\textwidth]{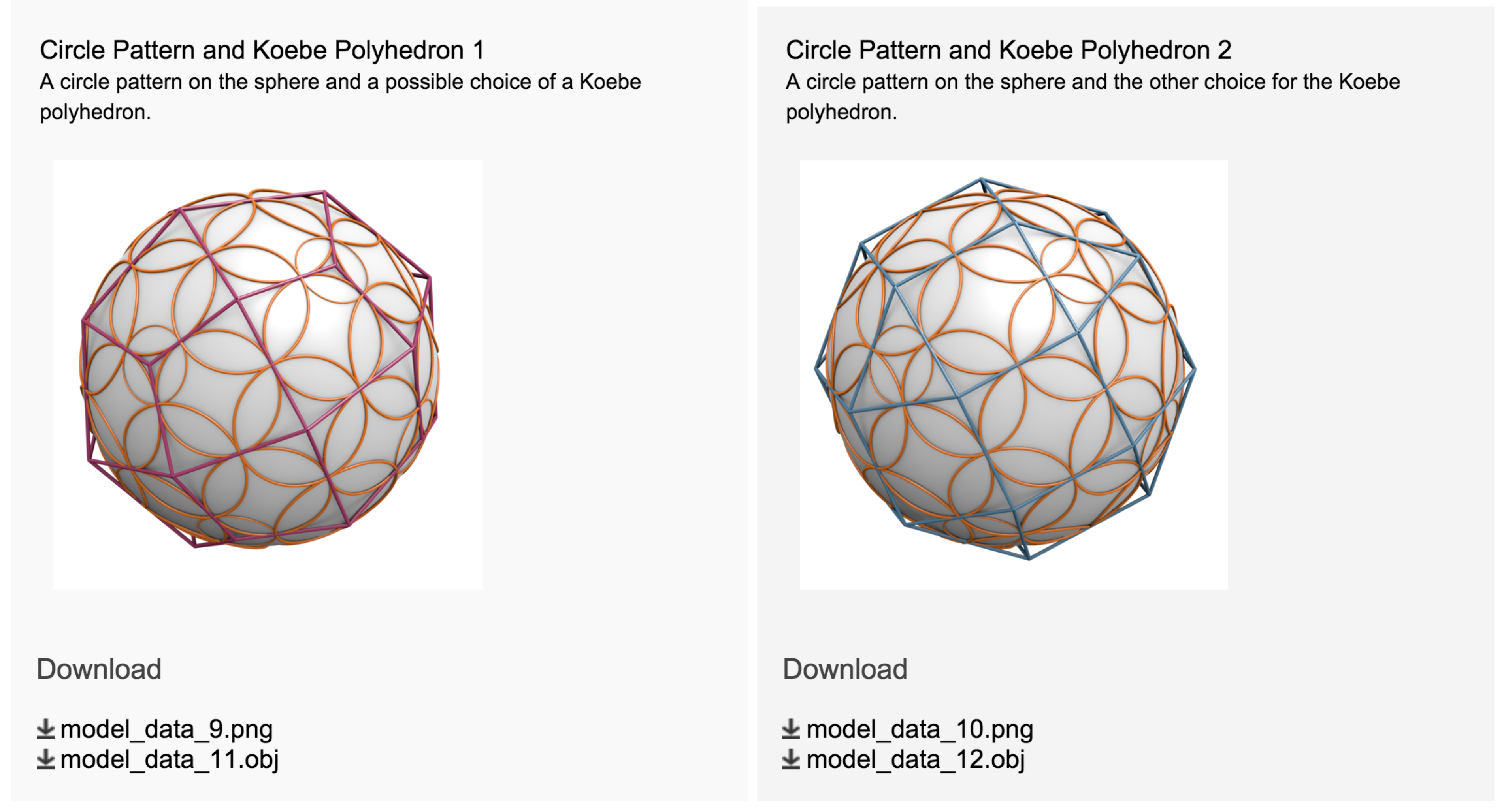}
\caption{Screenshot of two media objects of the "Koebe Polyhedra" model. 
The two images show the two corresponding Koebe polyhedra for a given circle pattern.}
\end{figure}

The first step for the construction of a Koebe polyhedron is to create an orthogonal circle pattern corresponding to the desired polytopal cell decomposition of the sphere.
This is generally done by finding critical points of a functional 
%(for full details see~\cite{BobHofSpr06}) 
expressed in the the variables $\rho_i=\log\tan\frac{r_i}{2}$ given by the spherical radii~$r_i$.
Once the radii are known the circles can be layed out.
The still remaining freedom of applying a M\"obius transformation can be fixed (up to a simple rotation) by requiring the center of mass to be at the sphere center.
The vertices of the circumscribed Koebe polyhedron that corresponds to the circle pattern can now easily be found by inverting the euclidean centers of the circles in $\mathbb{S}^2$ (the cone tips are the points polar to the planes containing the circles).
Here we have the freedom to choose one of the two orthogonal families of circles to become vertices of the Koebe polyhedron, and the other family to become faces.

The online model features a selection of Koebe polyhedra.
Each one with an OBJ geometry file and a PNG image file.

\goodbreak \subsection{Lawson's Surface Uniformization}
% MJ: "We" is not possible since the authors of the model do not agree with the authors of this paper;
% the other interpretation of "we" (author/s and reader combined does not make sense either).

Authors: Stefan Sechelmann, Alexander Bobenko, and Boris Springborn

\begin{quote}
\texttt{\scriptsize http://gallery.discretization.de/models/lawsons\_surface\_uniformization}
\end{quote}

%Fuchsian uniformizations of the Riemann surface of Lawson's genus 2 minimal surface in $\mathbb{S}^3$ \cite{Law1970} are presented in this model. The results were created using the discrete theory  as presented in \cite{BobSechSpr}.
%The model contains three different conformally equivalent representations of the surface and presents images of the corresponding hyperbolic tilings.
%% Namely a stereographically projected embedded version.
%% The model has been provided by Konrad Polthier~\cite{polthier97}.
%% The conformally equivalent algebraic curve $\mu^2=\lambda^6-1$.
%% And a conformally equivalent square-tiled surface glued from six squares.
%
%Lawson's surface is conformally equivalent to the algebraic curve $\mu^2=\lambda^6-1$, which is hyperelliptic.
%The branch points $\lambda_1,\ldots,\lambda_6$ are the 6th roots of unity. The model contains two different uniformizations of branched covers.
%
%An embedding of Lawson's surface in $\mathbb R^3$ is obtained via stereographic projection.
%The presented version realizes the hyperelliptic involution of the Riemann surface as a rotational symmetry.
%Its symmetry axis meets the surface in six points.
%These fix points of the hyperelliptic involution correspond to the branch points of the hyperelliptic algebraic curve. 
%
%Lawson's surface is conformally equivalent to a surface made of squares identified along suitable edges.
%A uniformization using the discrete metric defined by the edge lengths of the squares and its diagonals is presented.

\noindent
Fuchsian uniformizations of the Riemann surface of Lawson's genus 2 minimal surface in $\mathbb S^3$ \cite{Law1970} are presented in this model. 
The results were created in \cite{BobSechSpr} using the discrete uniformization theory. 
Three different conformally equivalent representations of the surface and of the corresponding hyperbolic tilings are presented.

Lawson's minimal surface in $\mathbb S^3$ is conformally equivalent to the hyperelliptic curve $\mu^2=\lambda^6-1$. The branch points $\lambda_1,\ldots,\lambda_6$ are the 6th roots of unity.

\begin{figure}[h]
\centering
\includegraphics[width=0.45\textwidth]{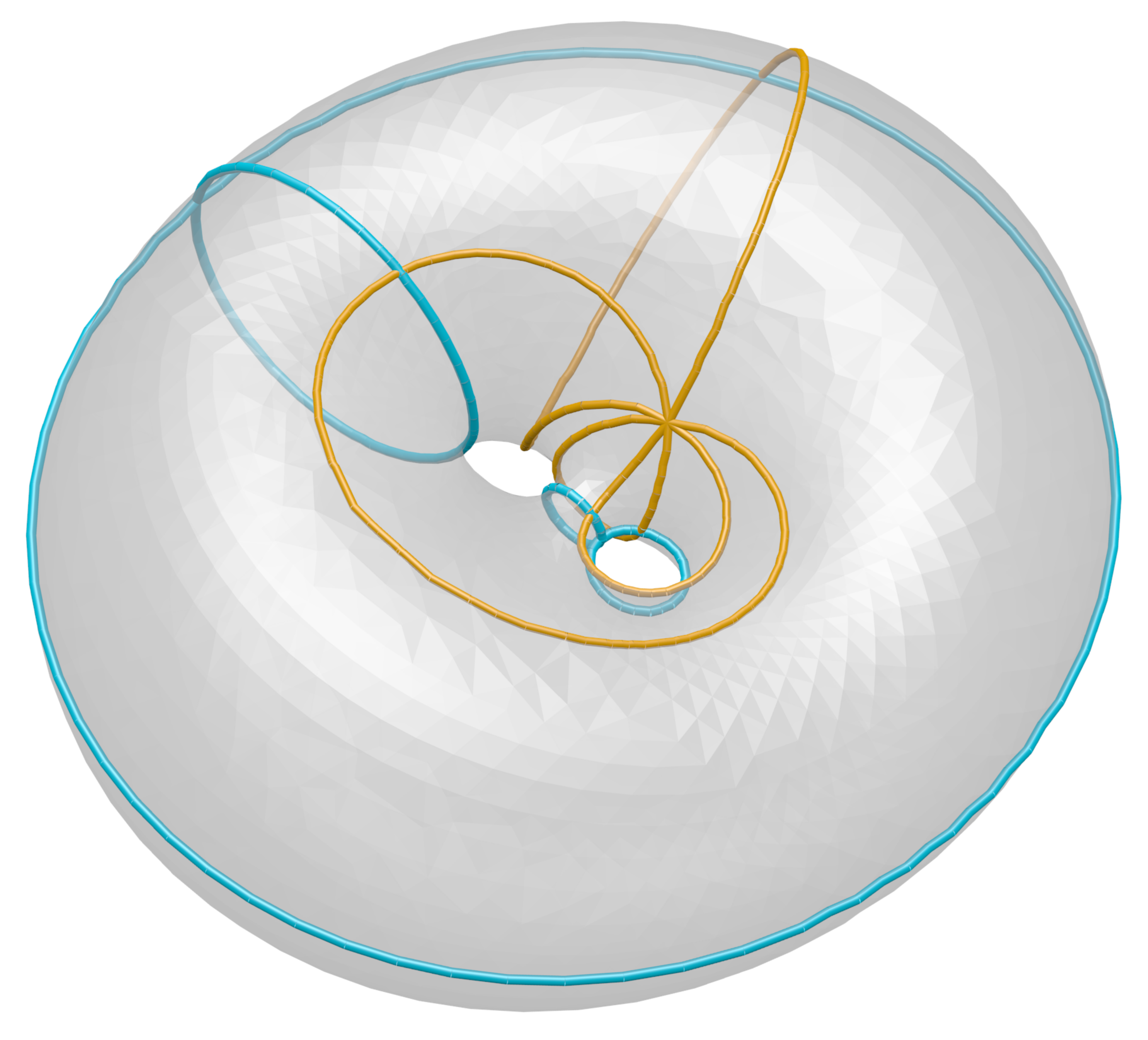}
\includegraphics[width=0.45\textwidth]{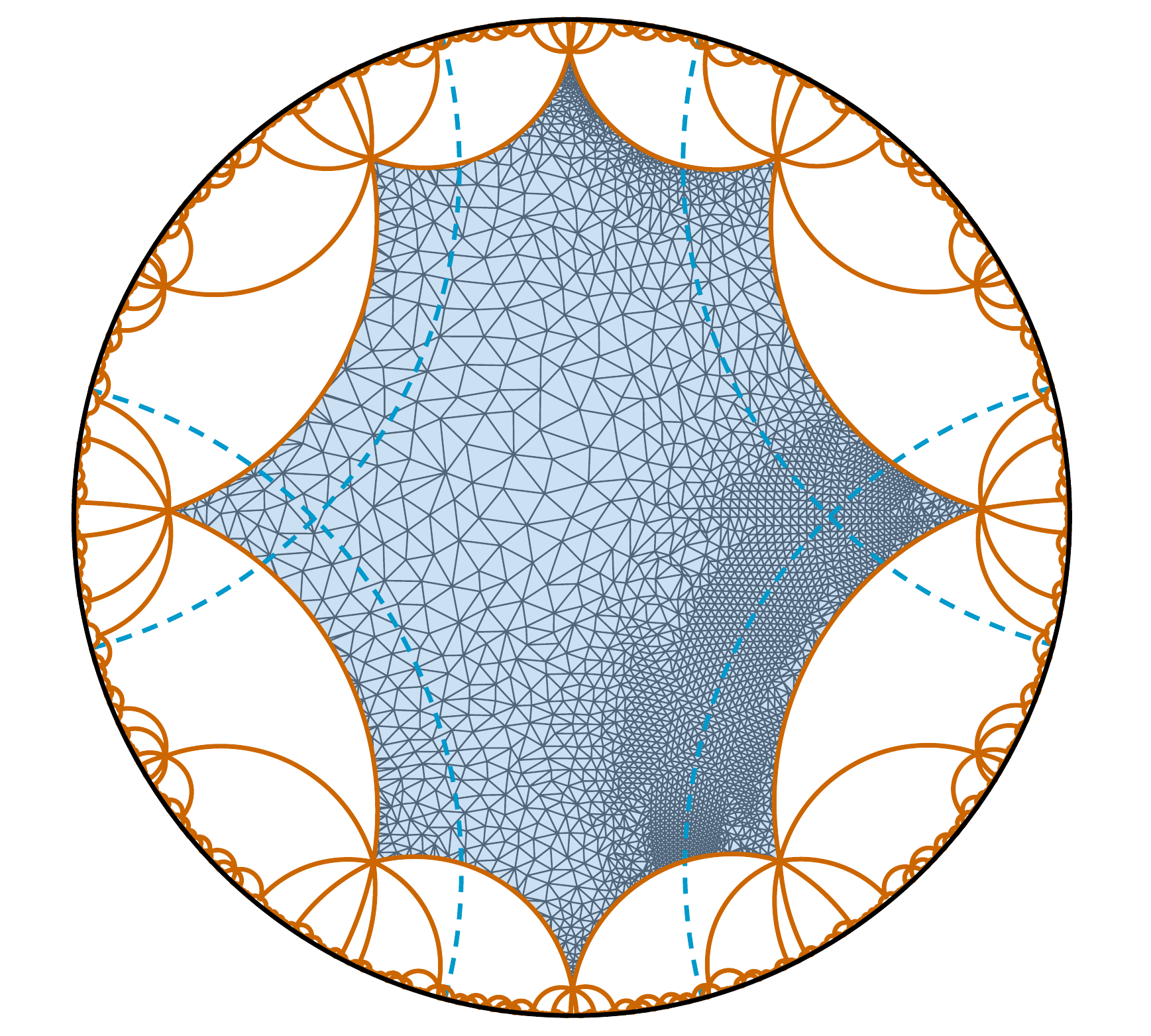} 
\caption{
Left: The Lawson surface in $\mathbb R^3$, the boundary curves of the fundamental domain of the uniformizing group in the right picture are shown in red. 
Blue curves correspond to simple closed geodesics corresponding to the axes of generators of the group. 
Right: The uniformization of the Lawson surface in the Poincar\'e model of hyperbolic space with a canonical fundamental domain (red) and axes of the hyperbolic generators of the uniformizing group.
}
\label{fig:lawson}
\end{figure}

An embedding of Lawson's surface in $\mathbb R^3$, see Figure~\ref{fig:lawson}, is obtained via stereographic projection from $\mathbb S^3$~\cite{polthier97}. 
For this surface the hyperelliptic involution of the Riemann surface is realized as a rotation by $180^\circ$. 
The axis meets the surface in six points, which are the branch points of the hyperelliptic curve.

The third realization of the Riemann surface is made of squares identified along suitable edges. 
The fundamental domain is identified with the two others.

The model features the data of the discrete uniformizations in XML format. 
It contains the combinatorial data, the coordinates of the points, and the uniformizing groups data. 
PDF vector graphics and PNG images provide 2D renderings of objects in 3D space.

\break 
\subsection{Tropical Grassmannian TropGr(2,6)} 
\label{ex:trop_gr}
Authors: Michael Joswig and Benjamin Schr\"oter

\begin{quote}
\texttt{\scriptsize http://gallery.discretization.de/models/tropical\_grassmannian\_gr26}
\end{quote}

\noindent
Tropical geometry studies piecewise linear images of classical algebraic varieties.
Many interesting properties remain visible in the tropicalization.
Additionally, this method reveals relations between geometry and optimization.
One outcome are combinatorial algorithms for dealing with classical objects.

The \emph{tropical Grassmannian}
TropGr$(d,n)$ is the tropicalization of the classical Grassmannian Gr$(d,n)$, defined over some field.
It parameterizes the tropical $d$-planes in the tropical $(n-1)$-torus; see \cite[\S4.3]{TropicalBook}. 
For $d=2$ the tropical Grassmannian coincides with the corresponding \emph{Dressian}, which arises as the subfan of the secondary fan of the hypersimplex $\Delta(d,n)$ corresponding to those regular decompositions whose cells are matroid polytopes \cite{HJJS:2009}.

\begin{figure}[h]
\centering
\includegraphics[width=\textwidth]{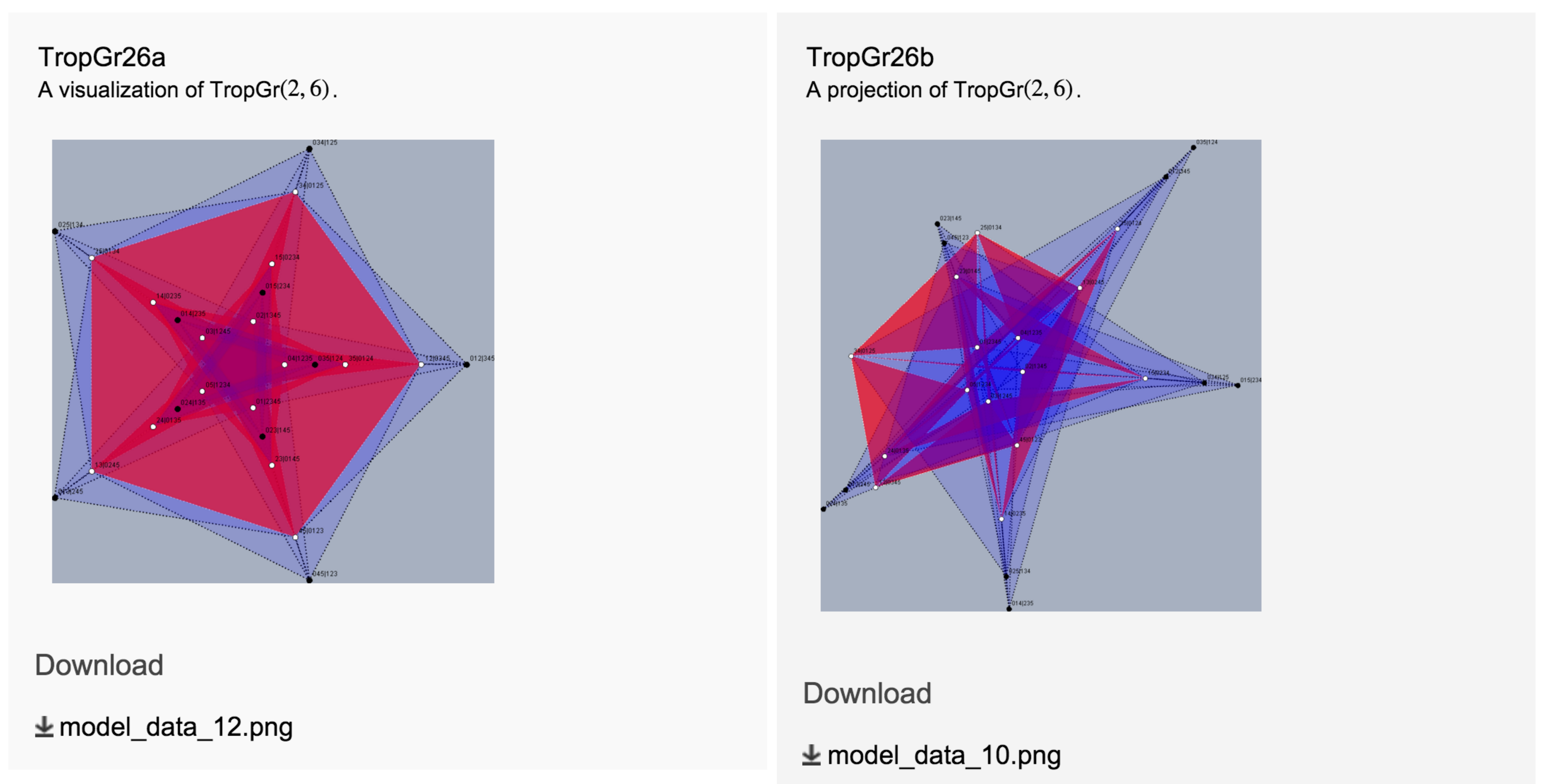}
\caption{Screenshots of two images contained in the "Tropical Grassmannian TropGr(2,6)" model.} 
\end{figure}

How to properly visualize TropGr$(2,6)$ is far from obvious, since (modulo its lineality space and intersected with the corresponding unit sphere) this is a $2$-dimensional spherical simplicial complex naturally embedded in the $8$-sphere.
It has $25$ vertices, $105$ edges and $105$ triangles.
The approach here employs a fixed copy of smaller tropical Grassmannian TropGr$(2,5)$, obtained by deletion, as a frame of reference and uses projections.
The \emph{deletion} of a matroid as a smaller matroid which is induced on fewer elements, and this notion carries over matroid decompositions.

The media objects associated with this model are a \texttt{polymake} \cite{DMV:polymake} description and pictures of various projections, in PNG format.
%\todo{JT: The picture does not show the polymake files yet.}
% There is nothing to show: it's an aweful XML file descibing a high-dimensional fan.  It will soon be added to the model.
% I was referring to a missing download link in the picture, not a visualization.
% But maybe it would be better to only show pictures after all, instead of screen shots of the download view.

\section{Architecture} 
\label{sec:architecture}

In this section we describe the structure of a model and the organization of data within the \dgdgallery.
It is also explained how users create, edit, and interact with models using model permissions.
Finally, we give details on the submission system and the review process.

\subsection{What Is a Model?}
\label{sec:model}

\begin{figure}[ht]
 \centering \resizebox{\textwidth}{!}{\includegraphics{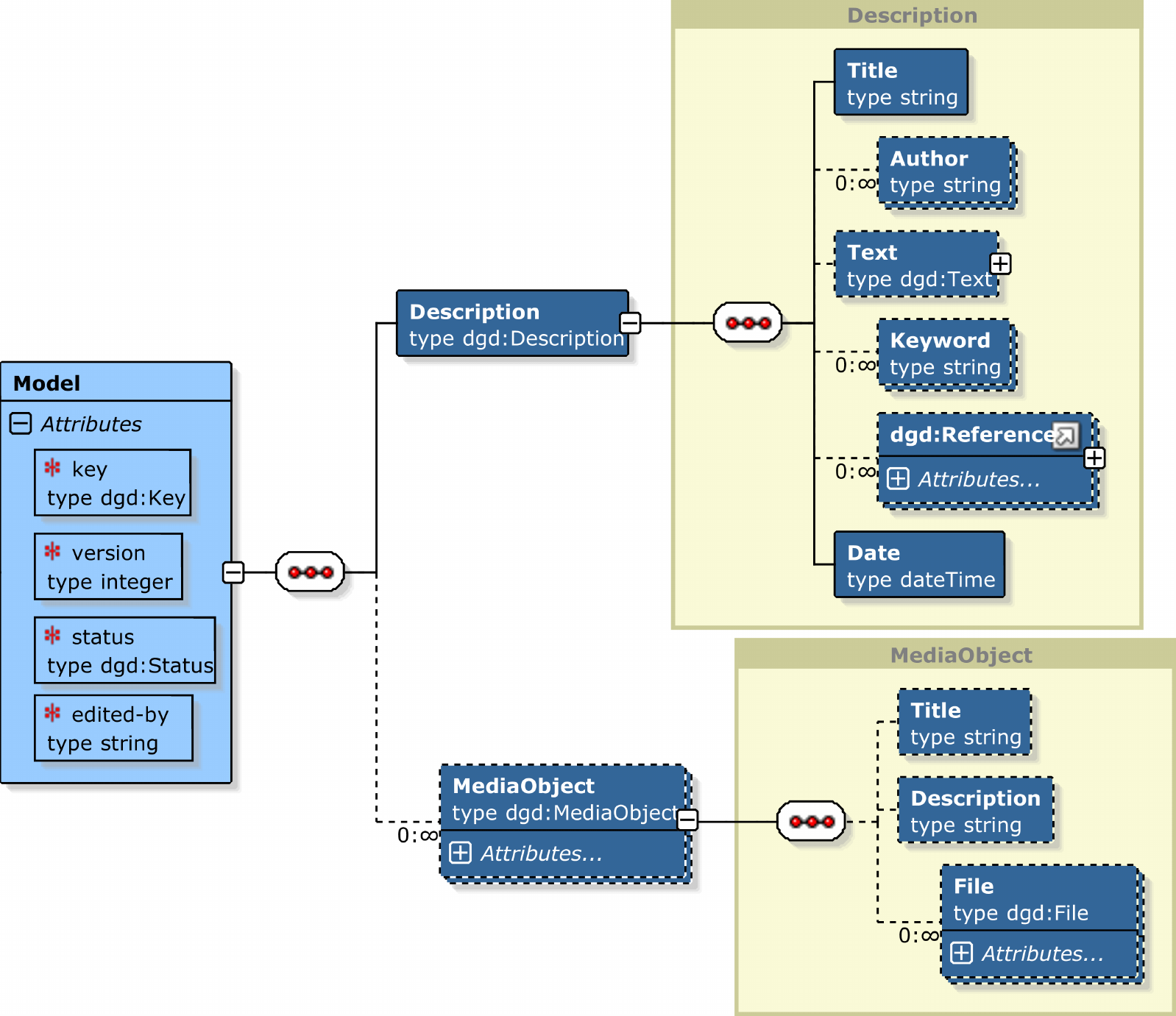}}
 \caption{The structure of a model.
  A model is a collection of media objects together with a description.}
 \label{fig:model_structure}
\end{figure}

The architecture of the \dgdgallery is built around the definition of the \texttt{Model}, see Figure~\ref{fig:model_structure}.
The \texttt{teletype} font is used to indicate that a word is the name of an abstract data type, one of its attributes, or an admissible value.
From a high level perspective a \emph{model} is a collection of files together with a description.
The description contains fields for the title, authors, a description text, keywords, literature references, and the creation date.
The data files associated with a model are bundled into media objects.
A \emph{media object} is a set of files together with a title and a description text.
% that's like a definition here
These files may be images, videos or data for specific software systems.
While some file formats are more common (and more reasonable) than others, conceptually we allow for any file format to become part of a media object.
In this way our design is very flexible and thus could be applied in other contexts.

The data type {\tt Model} has a {\tt key}, a {\tt version} number, a {\tt status} field, and an {\tt edited-by} username.
The model key is a unique identifier that is used, e.g., to assemble the permanent link of the model on the web.
The version number is assigned automatically for keeping track of a model's history.
Throughout the following the word ``model'' refers both to a specific version and to the entire history of a model.
The standard representative of a model is given by its latest version.
The {\tt edited-by} field contains the username of the author of a particular model version.
Hence, in a database a model can be uniquely identified by its key, a model version is identified by its key and version number.

The status of a model can take the values {\tt edit}, {\tt pending}, {\tt rejected}, or {\tt approved}.
See Section~\ref{sec:submission_system} for a detailed description of the model status and the submission process for models in the gallery.

The model description is a collection of the following information that is provided by the editor of a model.
While this somehow resembles the structure of a traditional research paper, there are some notable differences.

\begin{itemize}
\item
 The {\tt Title} of the model.
\item
 A sorted list of {\tt Author}s.
 Since this is a frequent source of misunderstanding, it is worth explaining.
 The authors of the model are those who create the content that is presented online.
 Like for a research paper all the scientific work that leads to the model must be properly acknowledged in the references.
 However, clearly, the set of authors cannot comprise all the authors who contributed something to the entire history of a mathematical idea.
 For instance, suppose that Alice first describes a new type of surface in a traditional research paper, and Bob afterwards produces a model from Alices description (without Alice's help).
 Then Bob is the only author, who must cite Alice's paper.
\item
 The description {\tt Text}, which can contain any valid {\LaTeX} source code that can be compiled using the {\sc MathJax} library, see~\cite{mathjax-website}.
 References to the literature or to media objects can be cited via the {\tt \textbackslash cite\{.\}} command.
 Previews of media objects can be included with the {\tt \textbackslash media\{.\}} command.
\item
 A set of keywords can be assigned to the model.
 The keywords are used on the web-site to, e.g., improve search features.
\item
 A set of references each of which consitsts of a reference key that is to be used in {\tt \textbackslash cite} commands and a set of key value pairs.
 The user interface of the gallery maps {\BibTeX} entries to model references.
 Conversely a model reference is rendered and referenced using common {\BibTeX} styles.
\item
 The date field of the description contains the creation date of the particular version of the model.
\end{itemize}

A model contains a number of media objects for visualization and use in other software systems.
This concept will be explained below.

\subsection{Media Objects and Data Files}
\label{sec:media_objects}
A media object is a collection of files that describe the same set of data associated with the model.
For instance, several media objects might correspond to various views of the same model; e.g., see the tropical Grassmanian in Example~\ref{ex:trop_gr}.
A different use case for several media objects for the same model is displayed for the discrete catenoid and helicoid model in Example~\ref{ex:catenoid}.
One media object shows a catenoid, whereas the other media object contains a dynamic rendering of the transformation from the catenoid to the helicoid.

The various file formats for one media object are meant to display one view of the model on several backends.
For instance, the discrete catenoid media object comes with a PNG file to be displayed in a standard web browser and with an OBJ file which allows 3-dimensional interactive visualization with a suitable viewer software. 
The data files comprising the media objects are stored in the file system separately from the model database.
The media objects of a model contain links to those data files, see Section~\ref{sec:model}.

In principle, we do not restrict the file formats for data files of any media objects.
This makes the \dgdgallery very flexible, but this also creates potential trouble with file formats that are uncommon.
We support the direct visualization of a few well chosen standard file formats. 
So far these include the following: PNG and JPG (for raster image data), SVG and PDF (for vector graphics), OBJ (for 3-dimensional geometric data), MOV, MP4, and OGV (for video content), POLY (for {\tt polymake} data), and others.
Interactive content is not excluded, see Section~\ref{sec:frontend}, but the danger of a particularly low stability over time should be well considered.
We rely on the review process for a sound selection.

\subsection{Versioning}

The \dgdgallery tracks the history of each model via the {\tt version} attribute, see Figure~\ref{fig:model_structure}.
Editing a model amounts to adding a new version with modified content.
If a model is deleted, all versions of the model and the data files linked are deleted from the database.
A data file is kept in the file system as long as there exists a link to it from some version of a model.

The version system is particularly useful for models with several authors who can collaborate through our front end.

It is worth noting that we also allow published models to be edited further and resubmitted.
Upon acceptance this new version will appear as the current version of the model on the web page.
The previously published versions remain visible and can be compared.
This way authors can keep their models up to date; see also Section~\ref{sec:migration} below which describes our migration process.

\subsection{Users}
\label{sec:users}

% \begin{figure}
% \centering
% \resizebox{0.5\textwidth}{!}{\includegraphics{gallery_User.pdf}}
% \caption{User structure. A user object contains login data for the web site as well as plain text name, email address, and permission information.}
% \label{fig:user_structure}
% \end{figure}

The users of the \dgdgallery are represented by their usernames, i.e., their {\tt login} names on the web-site.
%, see Figure~\ref{fig:user_structure}.
The access is password restricted, see also Section~\ref{sec:backend}.
In addition to the username and password we store the name and email address of the person that is associated with the user.

\begin{figure}[ht]% positioning right after section heading looks bad; also: technically belongs to user section
 \centering \resizebox{\textwidth}{!}{\includegraphics{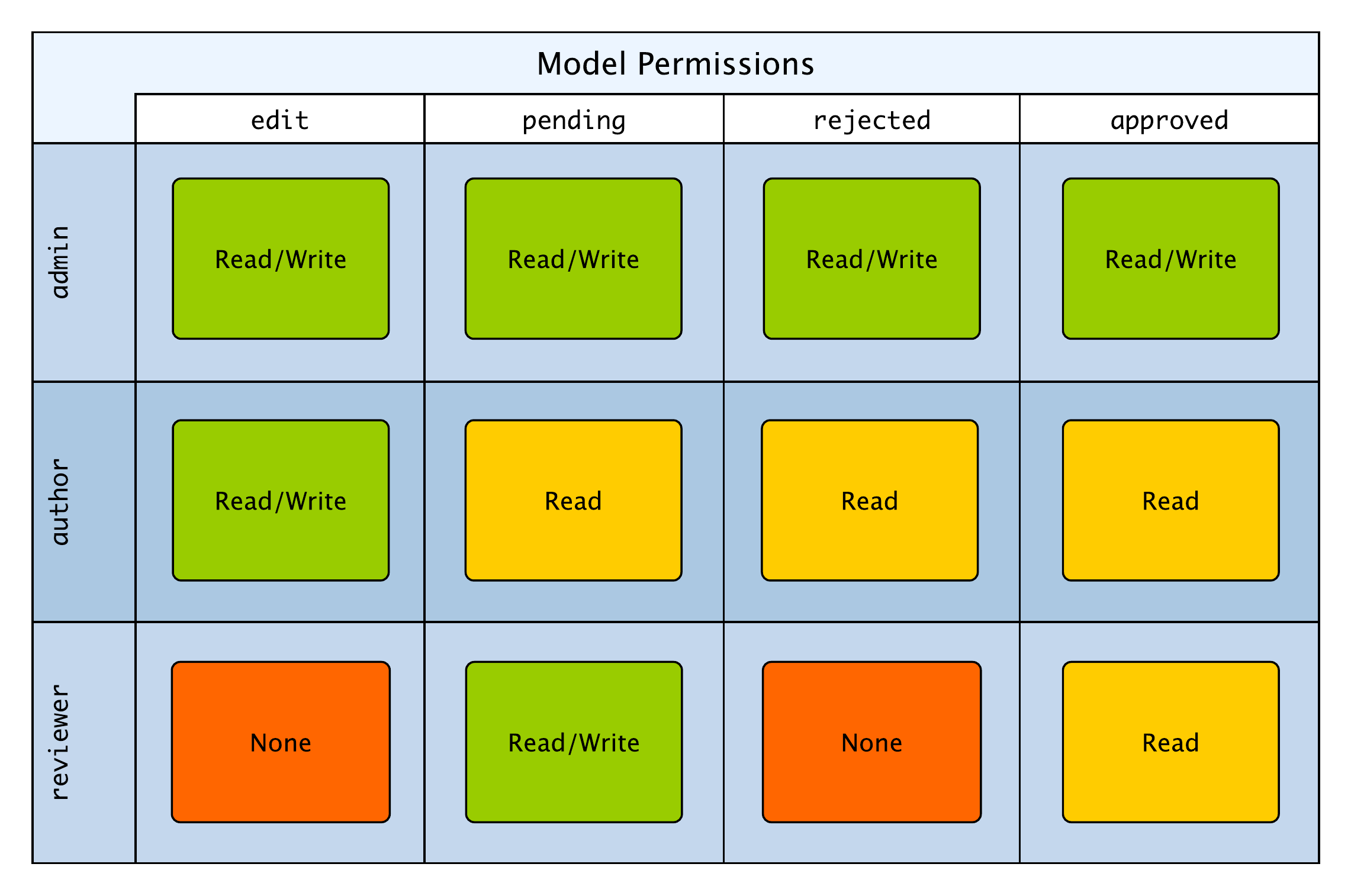}}
 \caption{
Read and write permissions during the life-cycle of a model.
A user with global {\tt admin} privileges can read and write on the model at any state (first row).
The {\tt author} of the model can edit his model if it is in {\tt edit} state (second row). 
A {\tt reviewer} can edit a model if it has been submitted ({\tt pending} state, third row).
}
 \label{fig:permissions}
\end{figure}

A user has a global user-role that can take the values {\tt admin}, {\tt reviewer}, or {\tt author}.
In addition to the global user-role we store model-roles for each model associated with a user.
A user can be the {\tt owner} or an {\tt editor} of a model.
The read and write access to models is restricted such that it is based on a combination of the global user-role, the model-role and the state of the model, Figure~\ref{fig:permissions}.
This implementation allows reviewers to act as model authors but prevents them from approving their own models.

A notable design decision is that a reviewer can modify a submitted model to correct obvious typos and other minor changes before approval.
Each owner of a model can invite other users to become either owners or editors of that model.

\subsection{Submission Process}
\label{sec:submission_system}

The \dgdgallery uses a submission system to publish models on the web-site.
The idea is that a board of reviewers approves, sends back for revision, or rejects a submitted model.
The review process should concern the quality of the content and address technical issues with the digital data.
The review board has to work out and agree on some quality criteria for a model.

During its life-cycle a model has assigned a status value.
A newly created model starts its life in {\tt edit} state.
It can be previewed and edited by the {\tt owner}s of the model, typically the creator of the model, and any additional user with the {\tt editor} model-role, see Section~\ref{sec:users}.

A model can be submitted by a user with the {\tt owner} model-role.
The status of the model changes to {\tt pending}.
A model with {\tt pending} status is read-only for the {\tt owner} and all {\tt editor}s.

Reviewers can preview and edit {\tt pending} models.
A reviewer edits a model to resolve small issues such as typos.
If the quality of a model is sufficiently high then a reviewer can accept a model.
The status of the model is changed to {\tt approved}.
If the content has flaws or technical issues that can be resolved by the creator of the model, the reviewer sends the model back to {\tt edit} state.
Any action by a reviewer is accompanied by a review text, which is presented to the authors of the model.

If a model is sent back for revision, the authors can edit the model according to the review text and resubmit.
If the model is {\tt rejected} it can neither be edited nor resubmitted.
A model will be rejected if it contains major flaws or its content is not appropriate for publication in the \dgdgallery.

Approved models become publicly available on the \dgdgallery web-site (see Section~\ref{sec:publication}).
To further improve public models, e.g.\ by correcting errors or replacing outdated file formats, a new version of an approved model can be created, which is back in {\tt edit} state.
To publish the new version it has to be submitted und undergo the revision process again.

The model submission system dispatches messages to the users of the \dgdgallery on every model status change.
Reviewers are notified about submitted models.
Model owners/editors are notified upon acceptance, rejection, or call for revision.

In principle any reviewer can accept, send back, or reject a model.
We rely on a reasonable communication between the reviewers to organize the review process.
Accepting a model is based on formal correctness, technical soundness, mathematical content and visualization quality.

\subsection{Publication and Licensing}
\label{sec:publication}

Content that has been approved by the board of reviewers is published on the \dgdgallery web page.
The presentation of the content on this page is equivalent to the preview during {\tt edit} state of the model.
The {\tt key} defines the permanent absolute URL of a model: 
\begin{quote}
{\url{https://gallery.discretization.de/model/<key>}}.
\end{quote}

\noindent
The content of the \dgdgallery is published under the Creative Commons Attribution-ShareAlike 4.0 International license, short CC BY-SA 4.0, see~\cite{cc-license}.
This means in particular that we allow for our data to be used commercially, enabling newspapers or commercial web blogs to include content from the gallery without further complications.
Appropriate credit must be given if any content is reproduced or used, and this includes a link to the \dgdgallery.

\section{Implementation}
\label{sec:implementation}

In this section we elaborate on the technical decisions that we made in order to implement the {\sc DGD gallery}.
It should give an impression of the system architecture, libraries, frameworks, and languages in use and their respective purposes.

\begin{figure}
 \centering \resizebox{\textwidth}{!}{\includegraphics{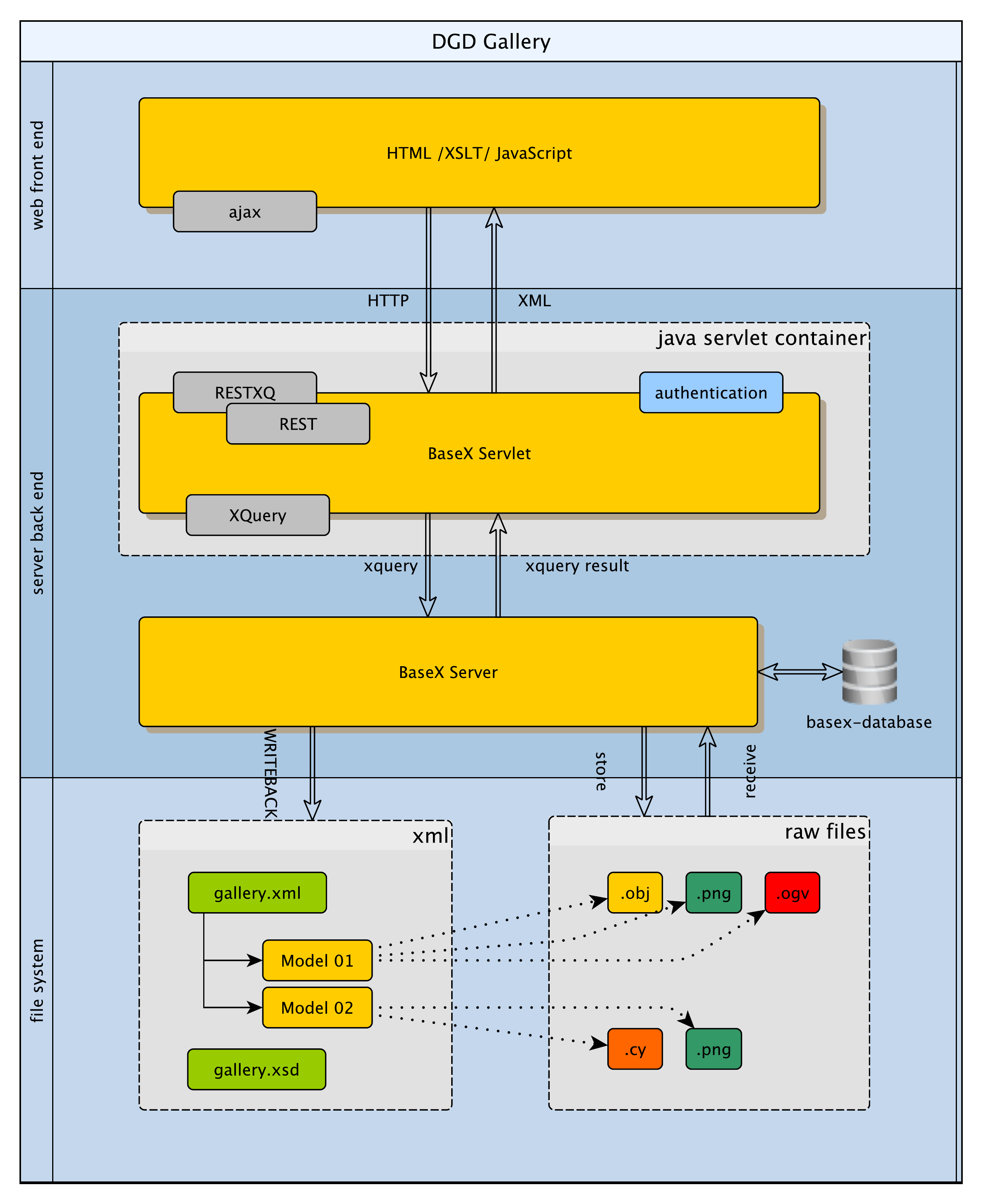}}
 \caption{ System architecture of the \dgdgallery.
  On the file system level we store XML model data and data files.
  A {\sc BaseX} server manages the read/write access to the XML documents and data files.
  It maintains an instance of a document database to optimize access to the XML data.
  At the same time a BaseX servlet provides a REST API \cite{rest-wiki-website} to connect the HTML/Javascript web front end of the gallery.
  It runs inside an Apache Tomcat servlet container executed within an Apache HTTP web server.
  The front end uses AJAX techniques and XSLT to create an interactive application using the API provided by the application server.
 }
 \label{fig:architecture}
\end{figure}

We imposed some a priori constraints on the implementation mainly to ensure reusability and persistence of the data over time.

\begin{itemize}
\item
 {\bf Human readable data format} (with enough structure to allow for easy validation and transformation): We chose XML for storage on the server and as the web server API data format.
 It fits the tree-like structure of our data and can be transformed to anything else, e.g.\ using XSLT. 
 This allows for easy migrations which can range from changing the structure of the models to getting rid of XML itself (replacing it with some more sophisticated data format in the future).
 To ensure that all stored data, and in particular data entered by users of the system, agrees with our specifications we use the XML Schema concept \cite{xmlschema-web}.
 This allows to validate all data on insert and during migration, see Section~\ref{sec:migration}.

\item
 {\bf Database framework agnostic storage} (while still using a database): The XML and any binary data are stored and handled using the XML document database {\sc BaseX}~\cite{basex-website}, see Section~\ref{sec:backend}.
 This gives us low access times (for data cached in main memory) and a transparent mechanism for permanent storage on the server's file system.
 The binary data files of the model's media objects are stored next to the XML data and linked appropriately, see the file system section in Figure~\ref{fig:architecture}.

\item
 {\bf Separation of data and presentation}: We separate our application into a back end (on the server for database management only) and a front end (creating a HTML representation on the client machine).
 The {\sc BaseX} database already provides the means for a complete implementation of the back end via {\sc XQuery}.
This includes the specification of a REST API \cite{rest-wiki-website}, which is a standard way to define the communication interface between server and client in the internet.
The API returns XML or binary data in response to specified HTTP requests from the front end.
 XML is already close to HTML, while still not carrying explicit information on the visualization.
 This allows for the easy generation of multiple presentations from just one XML. In Section~\ref{sec:frontend} we elaborate on the front end, which is based an XSLT, JavaScript and AJAX \cite{ajax-wiki-page}.
% \item {\bf Use of well established frameworks}:
%   In the context of XML data storage and processing {\sc BaseX} is by now well-established, see~\cite{basex-website}.
% BaseX was mentioned often enough
\end{itemize}

\subsection{XML Based Backend and the XML Document Database {\sc BaseX}}
\label{sec:backend}

We use the established XML document database {\sc BaseX} for storing our data.
This automatically provides us with permissions, versioning, and life-cycle management for the models.
{\sc BaseX} runs on any Java application server.
We use Apache Tomcat~7, see~\cite{tomcat-web}.

{\sc BaseX} allows for the implementation of (web) applications using the XML query language {\sc XQuery}, see~\cite{xquery-web}.
It combines the database access and application server logic implementation into one language.
Additionally {\sc BaseX} can be used to implement a REST API via RESTXQ, which is a set of {\sc XQuery} annotations for handling HTTP requests and generating HTTP responses \cite{restxq-docs}, see Listing~\ref{alg:create_model_api}.

\begin{lstlisting}[ 	caption={
XQuery with RESTXQ annotations.
The function api:createModel defines the web API function to create a model for a specified user and title string.
The RESTXQ annotations, lines~8--13, define the REST interface of the server.
User permissions are checked, line~19, and the corresponding database function to create a new model is called, line~20.
User credentials ({\tt\$user}, {\tt\$pass}) are optional parameters and are transmitted using HTTPS API calls. Once logged in we use session cookies to authenticate users.
 }, 	label=alg:create_model_api, 	language=XQuery, 	captionpos=b, 	numbers=left ]
(:~
: REST API function to create a new model.
:
: @param $title mapped to POST parameter title, the title of the new model
: @param $user optional user name if no session can be inferred by the server
: @param $pass optional password
:)
declare %rest:POST
%rest:path("/createmodel")
%rest:form-param("title", "{$title}")
%rest:form-param("user", "{$user}", "dummy_id")
%rest:form-param("pass", "{$pass}", "")
%output:method("text")
%updating function api:createModel(
    $title as xs:string, 
    $user as xs:ID, 
    $pass as xs:string
) {
    let $user := user:checkUser($user, $pass)
    return model:createModel($user, $title)
};
\end{lstlisting}

Generally, all API calls to our back end have to be authenticated.
Either a username and password pair, or a session-cookie has to be provided along each request.
The front end implementation uses session-cookies, which are obtained by an authenticated call to the login API function.
A user's password is stored in the form of a salted bcrypt hash to provide protection against password recovery through an attacker in case of a server breach, see~\cite{Provos99}.

\subsection{A Fail-safe Release and Migration Process}
\label{sec:migration}

While a project like the \dgdgallery evolves the precise technical requirements for the database are likely to change.
This means that old versions will have to be migrated into new ones.
We implemented a release process for new versions of the web application and its data using Apache ant~\cite{apache_ant}.
We use XSLT 2.0 and XML Schema to define and validate database migrations~\cite{xslt2-web}.

In principle this allows for more general migrations than just XML to XML conversions between different schema versions of the database.
We can envision scenarios in the future where XML may turn into a legacy format and will be replaced by a more general versatile format.
With XSLT we can also convert our XML data into arbitrary text based formats allowing for a final conversion into file formats entirely different from XML.

\subsection{A JavaScript Web Front End}
\label{sec:frontend}

The standard way to enter a new model into the \dgdgallery is through our web front end.
This part of the application is completely separated from the back end, relying only on the REST API to \textsc{BaseX} for communication.

We use the AJAX scheme of web application development.
The application, with its HTML, XSLT, and JavaScript components, is initially loaded from the server.
The access to the database is organized as HTTP connections via JavaScript.
Once XML model data has arrived from the server we process it with XSL Transformations~\cite{xslt2-web} to provide dynamic HTML and JavaScript for each client.
We use the SaxonCE XSLT JavaScript framework to execute XSLT 2.0 in the browser, see~\cite{saxon-web}.

Media renderers are provided for several common media formats, see Section~\ref{sec:media_objects}.
In the case of images and videos we are relying on the standards built into HTML5.
We have support for the web capabilities of Cinderella to allow for interactive content~\cite{cinderella-website, richter2012cinderella}. For the web browser in particular we use CindyJS \cite{cindyjs-github-page}, an open source JavaScript variant of Cinderella that aims to be compatible with Cinderella.

% \section{Discussion and further development}

% \todo{JT: Separate file manager?}
% \todo{JT: Update possibility?}
% \todo{JT: Parametrized geometry? geometry generating scripts...}
% \todo{JT: Mathematics Subject Classification?}
% \todo{Symbolic calculation renderer?}
% We are happy with the reliability of the system. 
% On the other hand the XSL transformations on the client machine are, especially on weak systems, the limiting factor of a smooth user experience. 
% We are considering to move more client code, especially the HTML generation, to the server to increase the speed of the application. 

% Considering compatibility with mobile devices and media rendering we are discussing to switch to from our own javascript implementation to client frameworks such as Bootstrap~\cite{bootstrap-webpage} and Angular.js~\cite{angularjs-webpage}.

% % too marginal!
% % We want to include AMS Mathematics Subject Classification codes either as a replacement or parallel to the keywords provided currently with models.

\section*{Acknowledgement}
This research was supported by DFG SFB/TR 109 ``Discretization in Geometry
and Dynamics''.

\bibliographystyle{plain} \bibliography{dgdgallery}

\end{document}